\documentstyle[11pt,aaspp4,flushrt]{article}
\slugcomment{Astronomical Journal, in press}

\lefthead{}
\righthead{}

\begin{document}

\title{Optical-IR Spectral Energy Distributions of $z>2$ Lyman Break Galaxies
\footnote{Based on observations made with the
NASA/ESA {\it Hubble Space Telescope} obtained at the Space Telescope
Science Institute, which is operated by the Association of
Universities for Research in Astronomy, Inc., under NASA contract NAS
5-26555.}
\footnote{Based on observations made at the Kitt Peak National Observatory,
National Optical Astronomy Observatories, which is operated by the
Association of Universities for Research in Astronomy, Inc.\ (AURA)
under cooperative agreement with the National Science Foundation.}
}

\author{Marcin Sawicki and H. K. C. Yee}
\affil{Department of Astronomy, University of Toronto, Toronto, 
Ontario M5S 3H8, Canada
\\ email: sawicki@astro.utoronto.ca, hyee@astro.utoronto.ca}

% The abstract environment prints out the receipt and acceptance dates
% if they are relevant for the journal style.  For the aasms style, they
% will print out as horizontal rules for the editorial staff to type
% on, so long as the author does not include \received and \accepted
% commands.  This should not be done, since \received and \accepted dates
% are not known to the author.

\begin{abstract}

Broadband optical and IR spectral energy distributions are determined
for spectroscopically confirmed $z>2$ Lyman break objects in the
Hubble Deep Field (HDF).  These photometric data are compared to
spectral synthesis models which take into account the effects of
metallicity and of internal reddening due to dust.  It is found that,
on average, Lyman break objects are shrouded in enough dust (typically
$E(B-V) \approx 0.3$) to suppress their UV fluxes by a factor of more
than 10.  The dust-corrected star formation rate in a typical HDF
Lyman break object is $\sim 60 h_{100}^{-2} M_\odot yr^{-1}$
($q_0=0.5$).  Furthermore, these objects are dominated by very young
($\lesssim 0.2$ Gyr, and a median of $\sim 25$ Myr) stellar
populations, suggesting that star formation at high redshift is
episodic rather than continuous.  Typically, these star formation
episodes produce $\sim 10^9 h_{100}^{-2} M_\odot$ of stars, or $\sim
\frac{1}{20}$ of the stellar mass of a present-day $L^*$ galaxy.

\end{abstract}

% The different journals have different requirements for keywords.  The
% keywords.apj file, found on aas.org in the pubs/aastex-misc directory, 
% contains a list of keywords used with the ApJ and Letters.  These are 
% usually assigned by the editor, but authors may include them in their 
% manuscripts if they wish. 

%\keywords{
%}

% That's it for the front matter.  On to the main body of the paper.
% We'll only put in tutorial remarks at the beginning of each section
% so you can see entire sections together.

% In the first two sections, you should notice the use of the LaTeX \cite
% command to identify citations.  The citations are tied to the
% reference list via symbolic KEYs.  We have chosen the first three
% characters of the first author's name plus the last two numeral of the
% year of publication.  The corresponding reference has a \bibitem
% command in the reference list below.
%
% Please see the AASTeX manual for a more complete discussion on how to make
% \cite-\bibitem work for you.   

\section{INTRODUCTION}

Recent observational studies of the high-redshift Universe have begun
to yield direct information on the nature of the population of
``normal'' galaxies at $z>2$.  Color-selected searches have started to
produce growing, spectroscopically-confirmed samples (e.g.,
\cite{ste96a}; \cite{ste96b}; \cite{pas96}; \cite{low97}; \cite{ste97}), 
while larger samples of objects identified solely on the basis of
multicolor photometry have been used to boost statistics (e.g.,
\cite{mad96}; Sawicki, Lin \& Yee 1997; \cite{con97}).  Whether
spectroscopically confirmed or not, such high-$z$ objects are often
called ``Lyman-break galaxies'' or ``U-band drop-outs''.  The names
reflect the fact that in these galaxies the UV flux is severly
attenuated by the combined effects of internal and intergalactic Lyman
line and continuum opacity.  Unlike serendipitously discovered single
objects, these new samples of high-$z$ galaxies are large enough to
allow the study of global properties of the $z>2$ galaxy {\it
population}, such as their number density, luminosity density,
luminosity function, and rates of star formation and metal ejection.

Despite this recent progress, there is much that still remains unclear.
Based on number densities, star formation rates, physical sizes,
morphologies and masses inferred from spectral line widths, Steidal
et al.~(1996a) and Giavalisco, Steidel, \& Macchetto (1996) argue
that Lyman break objects are the spheroidal components of present-day
luminous galaxies, seen in the act of formation.
Using diagnostics similar to those employed by Steidel et al.~and
Giavalisco et al., Lowental et al.~(1997) favor the interpretation
that Lyman break objects are either progenitors of present-day
low-mass spheroidal galaxies or are starbursting building blocks
of more massive galaxies of today (see also Colley et al.~1997).

Estimates of star formation rates in Lyman break galaxies are based 
on their rest-frame UV fluxes (e.g., \cite{ste96a}; \cite{mad96}; 
\cite{low97}).  However, even moderate amounts of dust may significantly 
suppress the UV flux and, hence, the inferred star formation rate.
Meurer et al.\ (1997) have studied the UV spectral indices of HDF
Lyman break galaxies and have deduced that, if ages of their stellar
populations are $\lesssim 0.3$ Gyr, reddening due to dust supresses
the UV flux fairly strongly (on average, a factor of 15 at 1600 \AA ).
Using the same spectral indices, but assuming older ages (1 Gyr),
Calzetti (1997) found the attenuation a factor of 3 lower than that
arrived at by Meurer et al.\ (1997).  These two different results,
obtained from the same data, reflect the fact that both dust and aging
have similar effects on the UV shape of the SED and that one can
mimick the effects of the other.  This dust-age degeneracy can be
lifted by the addition of IR (i.e., rest-optical) data, which are
sensitive to age but much less so to dust.

Ages of star-forming objects at $z > 2$ are not well known.  Steidel
et al.\ (1996a) feel that ages younger than a few $\times 10^7 yr$ are
unlikely, arguing that such young ages would imply that a large
number of galaxies have to be bursting simultaneously.  However, apart
from the gravitationally lensed $z=2.72$ OB star absorption-line
galaxy cB58 (\cite{yee96}), for which fits to broadband optical and IR
colors give the age of the dominant stellar population to be $\lesssim
35$ Myr (\cite{ell96}), no quantitative age measurements exist for
Lyman break galaxies.
% and we are reduced to argueing that the
%existence of galaxies at even higer redshifts ($z>4$), such as the few
%cases observed by, for example, Trager et al.\ (1997) and Franx et
%al.\ (1997), must mean that some relatively old ($\sim1$ Gyr) objects
%are in place by $z\sim2.5$.
%In view of the
%dust-age degeneracy, as well as for the dating of their evolutionary
%history, it is important to obtain constraints on the ages of the
%Lyman break galaxies.
Constraints on ages will lift the dust-age degeneracy, in addition to
providing clues to the evolutionary history of Lyman break objects.

The Hubble Deep Field (HDF;\cite{wil96}) is a small but intensly
studied area of the sky, for which both extremely deep multiwavelength
imaging (e.g., \cite{wil96}; \cite{dic97}; \cite{hog97}; \cite{ser97};
\cite{fom97}) and follow-up spectroscopy (e.g., \cite{ste96b};
\cite{coh96}; \cite{low97}; \cite{zep97}) have been carried out. 
To date, 17 spectroscopically confirmed $z>2$ Lyman break galaxies
have been reported in the HDF (Steidel et al.~1996a; Lowenthal et
al.~1997).

In this paper we construct the broadband optical and IR (i.e.,
rest-frame UV and optical) spectral energy distributions (SEDs) of the
17 spectroscopically confirmed Hubble Deep Field Lyman break galaxies
(\S~\ref{data}).  These observed SEDs are then fitted with
dust-corrected spectral synthesis models described in
\S~\ref{models_described}.  These fits show that light from Lyman break 
galaxies is attenuated by significant amounts of dust (typically
$E(B-V) \approx 0.3$; \S~\ref{fit_dust_content}) and is dominated by
very young ($\lesssim 0.2 Gyr$) stellar populations
(\S~\ref{fit_ages}).  Furthermore, Lyman break objects have high rates
of star formation (typically $\sim 60 h_{100}^{-2} M_\odot yr^{-1}$
for $q_0=0.5$;
\S~\ref{SFRs}), and quite low stellar masses (typically, $\sim 10^9
h_{100}^{-2} M_\odot$; \S~\ref{masses}).  Since internal reddening in
Lyman break galaxies appears to be significant, in
\S~\ref{discuss_SFR} we calculate its effects on the
recently-published estimates of $z>2$ star formation and metal
ejection rates.  In \S~\ref{discuss_ages} we speculate on the nature
of Lyman break galaxies in view of the high star formation rates and
young ages of stellar populations which we found earlier.  Our results
are summarized in \S~\ref{conclusions}.

%In \S~ref{models} the observed SEDs are then 
%fitted with spectral synthesis models 
%We do so in order to determine
%simultaneously the galaxies' intrinsic reddening content and ages of
%their stellar populations.  We find that in most objects moderate
%amounts of dust (typically $E(B-V) \approx 0.3$) are required, and
%that their stellar populations are consistent with being very young
%($\lesssim 0.2$ Gyr).  We also find that they appear to have high
%rates of star formation ($\sim 60 h_{100}^{-2} M_\odot yr^{-1}$ for
%$q_0=0.5$), and quite low stellar masses ($\sim 10^9 h_{100}^{-2}
%M_\odot$).

%The data used
%are described in \S~\ref{data}.  The photometric data are then
%compared with spectral synthesis models in \S~\ref{models}.  The
%implications of the observed ages and dust content are discussed in
%\S~\ref{implications} and the main results are summarized in
%\S~\ref{conclusions}.

\section{DATA}\label{data}

\begin{figure}
\plotone{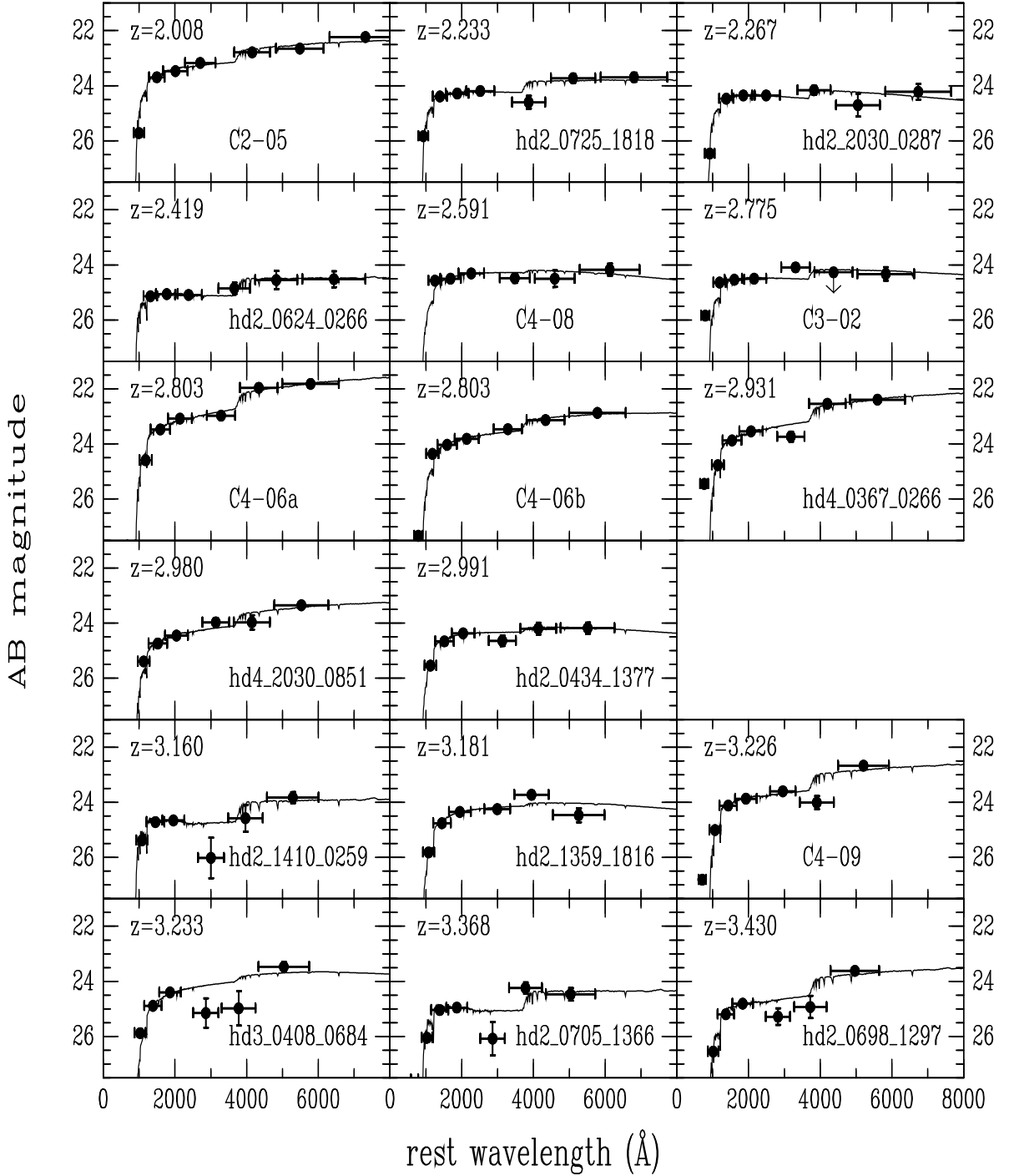}
\figcaption[Sawicki.fig1.ps]
{ \label{best_fit_mosaic.fig} The best-fitting $0.2Z_{\odot}$,
constant SFR models.  Dust content and age are free parameters.
Although their fluxes are plotted, the $U_{300}$ and $B_{450}$ data
were not used in the fit (see text).  The model SEDs shown include the
high-$z$ Lyman suppression of the $UV$ flux.  The bottom two rows
contain objects at $z>3$ --- objects for which fit quality suffers
from the fact that only one filter is present redward the Balmer
break.  Instantaneous burst and $0.02Z_{\odot}$ and $1.0Z_{\odot}$
models produced fits of similar quality as the ones shown here. }
\end{figure}

We study the broadband spectral energy distributions for the 17
spectroscopically confirmed $z>2$ HDF objects reported by Steidel et
al.\ (1996b; 6 objects\footnote{Steidel et al.\ report five $z>2$
redshifts.  Since, morphologically, C4-06 appears to have two distinct
components, we have treated it as two objects at a common redshift.},
some with revised redshifts\footnote{ The redshift of the Steidel et
al.\ (1996b) object C5-02 has been revised from $z=2.845$ to $z=2.008$
(Steidel, private communication).  The reported redshift of C3-02 may
also be erroneous: The observed broadband SED of C3-02
(Figure~\ref{best_fit_mosaic.fig}) exhibits a break between the
$I_{814}$ and $J$ filters and not, as would have been expected of the
Balmer break in a $z=2.775$ object, redward of $J$.  We note that the
absorption lines of \ion{C}{4} 1550, \ion{Fe}{2} 1608, and \ion{Al}{2}
1670 could be misidentified as Ly~$\alpha$ 1216, \ion{Si}{2} 1260, and
\ion{O}{1} 1303.  In this case the redshift of C3-02 would be
$z=1.95$, which is consistent with the photometric redshift
$z_{phot}=2.1$ of Sawicki et al.\ (1997) and with the placement of the
Balmer break between the $I_{814}$ and $J$ filters.  While adopting
the revised redshift for C5-02, we retain the published value for
C3-02.  We tested the effects of this possible redshift
misidentification on our analysis and found them to be small.}) and by
Lowenthal et al.\ (1997; 11 objects).  We measured broadband fluxes
using both optical (\cite{wil96}) and infrared (\cite{dic97}) data.
Specifically, in the optical we used the publicly-available Version 2,
$4\times4$ binned, HST images of the HDF; in the IR we employed the,
also publicly available, Version~1 Kitt Peak~4m IRIM observations.
The use of the IR data is crucial, as at $z\sim 3$ the optical data
alone give coverage only bluewards of rest-2000~\AA, while the IR
extends this to rest-5500~\AA, thereby spanning the age-sensitive
Balmer break.  Altogether, the available photometry spans 3000 --
22000~\AA\ (or 750 -- 5500~\AA\ at $z=3$) in 7 bandpasses ($U_{300}$,
$B_{450}$, $V_{606}$, $I_{814}$, $J$, $H$, $K_s$)\footnote{$U_{300}$,
$B_{450}$, $V_{606}$, and $I_{814}$ are used to denote the HST filters
$F300W$, $F450W$, $F606W$, and $F814W$.}, although, as will be
explained later, we do not include the $U_{300}$ and $B_{450}$ data in
our fits.

We smoothed the HST images to match the poorer, ground-based point
spread function (PSF) of the IR images. We then performed photometry
using the PPP faint galaxy photometry package (\cite{yee91}).  Colors
were measured in ``color apertures'': for each object, a color
aperture was chosen so as to maximize the S/N.  Photometric
calibration was based on the zeropoints provided in the STScI-HDF and
KPNO-IRIM-HDF web pages (\cite{fer96}; \cite{dic96}), while conversion
to AB magnitudes (for the IR data) was done using the $F_{\nu}$(0 mag)
fluxes of Wemstaker (1981).  The photometry results are presented in
Table~\ref{tbl-1}. The magnitudes listed are ``total magnitudes''
obtained by correcting the object's magnitude within the color
aperture by its growth curve in the $V_{606}$ image.  A typical total
magnitude aperture is 3.2\arcsec.

\section{COMPARISON WITH SPECTRAL SYNTHESIS MODELS}\label{models}

\subsection{The Models}\label{models_described}

For the analysis of the stellar populations of Lyman break galaxies,
we generated a set of model broadband SEDs constructed in the
following way: We started with the Bruzual \& Charlot (1996)
multi-metallicity spectral synthesis models; specifically, we used the
KL96 theoretical stellar spectra\footnote{In general, substantial
differences still exist between composite SEDs based on theoretical
and empirical stellar spectra.  We have, however, tested our dust and
age results with Bruzual \& Charlot solar-metallicity empirical models
and found that the differences were small for our purposes.}  with the
Salpeter $0.1 < \frac{m}{M_\odot} < 125$ IMF.  These Bruzual \&
Charlot models were attenuated with Calzetti's (1997; see also
Calzetti et al.\ 1994) empirical reddening recipe for star-forming
galaxies\footnote{Calzetti (1997) gives:
\[F_{obs}(\lambda)=F_0(\lambda)10^{-0.4E(B-V)k(\lambda)},\] 
where
\[ k(\lambda)= \left\{ \begin{array}{ll}
[(1.86-0.48/ \lambda)/ \lambda -0.1]/ \lambda +1.73 & \mbox{for $0.63
\micron \leq \lambda \leq 1.0 \micron$}\\ 2.656(-2.156+1.509/ \lambda -
0.198/ \lambda^2 +0.011/ \lambda^3) + 4.88 & \mbox{for $0.12 \micron
\leq \lambda < 0.63 \micron$}
\end{array}
\right.  \]}~, covering a range of $E(B-V)$ values.
The reddened SEDs were appropriately redshifted and then convolved
with the instrumental transmission curves (in the case of the HST
data) or filter transmission curves (for the Kitt Peak IR
observations) producing a suite of model colors.  For each object this
suite of model colors contains two possible star formation histories
(instantaneous burst and constant star formation), 221 ages (0--20
Gyr), three different metallicities ($0.02Z_{\odot}$, $0.2Z_{\odot}$,
and $1.0Z_{\odot}$), and a range of reddening ($E(B-V)=0.0$ to $0.5$
in steps of 0.02).

Why were these particular ingredients chosen for the models?  {\it
Initial mass function:} The initial mass function, at least in
low-redshift starbursts, appears to be independent of environment and
is consistent with the Salpeter IMF with a high ($\sim 100 M_\odot$)
upper mass cutoff (\cite{sta96}; see also \cite{mas95}).  We have thus
chosen to use the Salpeter $0.1 <\frac{m}{M_\odot} < 125$ IMF
available in the Bruzual \& Charlot (1996) library.  {\it Reddening
curve:} Since, by virtue of their color selection criteria
(\cite{ste96b}; \cite{low97}), the 17 spectroscopically confirmed HDF
Lyman break objects are star-forming, the use of the Calzetti (1997)
reddening curve for star-forming regions is appropriate.  Unlike the
commonly-used LMC and SMC reddening curves (\cite{fit86};
\cite{bou85}) which are are derived for stars alone, the Calzetti
curve holds for star-forming {\it regions}; it thus automatically
includes such effects as back-scattering and geometrical distribution
of dust and is therefore much more appropriate for correcting the
effects of dust on the photometry of star-forming galaxies
(\cite{gor97}).  Using the LMC reddening law would produce extinction
values 2.5--3 times lower than that which we find (\S~\ref{discuss_SFR})
with the more appropriate Calzetti extinction curve.  {\it
Metallicity:} Damped Ly-$\alpha$ systems at $z=2-3$ have metallicities
$Z \lesssim 0.1 Z_\odot$ (\cite{pet94};
\cite{lu96}; \cite{pet97a}); thus $0.2Z_{\odot}$ or $0.02Z_{\odot}$ are
reasonable values to expect for the metallicity of Lyman break
galaxies, while $1.0Z_\odot$ provides a useful check in case the
metallicity of Lyman break objects were significantly higher than that
of damped Ly-$\alpha$ clouds.  {\it Ages:} At $z \approx 3$ the
Universe is only 1 -- 2 Gyr old and no object within it can be older
than that; we have, however, allowed for model ages of up to 20 Gyr as
a consistency check of the fits.  {\it Star formation history:} In
reality, the most likely star-formation scenario is that of a burst in
which the star formation rate (SFR) declines with time.  The two star
formation histories used (instantaneous burst and constant SFR) thus
provide the bracketing cases of the likely star formation histories of
these objects.
%Burst
%ages and dust contents obtained from these two extremes are then
%likely to bracket the actual values of age and reddening.
Dust contents and ages of stellar populations obtained from these two
extremes will then bracket the actual values of age and reddening.

For high-redshift objects, the neutral hydrogen contained in
Ly-$\alpha$ clouds along the line of sight provides an extra source of
opacity in the UV (e.g., \cite{mad95}).  The amount of extinction
provided by this mechanism is stochastic and depends on the numbers,
redshift distribution, and column densities of the Ly-$\alpha$ clouds
along the line of sight to the target galaxy.  The galaxy's UV flux is
affected blueward of rest-1216~\AA\ and is almost completely
extinguished below rest-912~\AA.  Whereas this effect is of great
usefulness in identifying high-$z$ galaxies, its stochastic nature
introduces uncertainties below $\sim$ 1200~\AA\ in the SED models for
individual galaxies.  These stochastic uncertainties, combined with
the fact that the dust reddening curve is not well known below
1200~\AA, led us to decide not to fit the two bluest filters ($U_{300}$
and $B_{450}$).  Thus all the fits presented in this paper have been
done using the 5 reddest filters ($V_{606}$, $I_{814}$, $J$, $H$, and
$K_s$).

Metallicity ($Z=0.02Z_\odot$, $0.2Z_\odot$, and $1.0Z_\odot$) and star
formation history (instantaneous burst and constant SFR) were fixed,
giving six scenarios.  In each of these six scenarios, age since the
onset of star formation and amount of dust 
%(parametrized through the
%$E(B-V)$ parameter) 
were free parameters.  The models and the data were compared by means
of $\chi^2$ minimization.  As an illustration,
Figure~\ref{best_fit_mosaic.fig} shows the fits thus obtained with the
$0.2Z_{\odot}$ con\-ti\-nu\-ous-SFR models.  Instantaneous burst
models, as well as those for $Z=0.02Z_{\odot}$ and $1.0Z_{\odot}$,
produced fits of similar quality.  Note that the two bluest filters,
although not used to produce the fit, match the best-fitting model
SEDs quite well.

Because reddening increases towards shorter wavelengths, the part of
the SED which contains information about dust attenuation is the
rest-frame UV.  Dust attenuation can, however, be mimicked by aging of
the stellar population: as the population ages, massive stars die out
and the UV flux decreases.  For example, below rest-3000\AA\, a
zero-age SED with $E(B-V)=0.3$ looks very much like a 100 Myr,
constant SFR, dust-free one.  The availability of IR data allows the
lifting of this degeneracy between age and reddening: at $z>2$, IR
data are probing the rest-optical part of the SED and consequently
measure the age by means of the size of the Balmer break and, to a
lesser degree, the slope of the continuum redward of
rest-4000\AA. Metallicity effects could pose a problem because of the
age-metallicity degeneracy (e.g.,
\cite{wor94}), but we have monitored their effects by performing fits
with models of different metallicity ($0.02Z_\odot$, $0.2Z_\odot$, and
$1.0Z_\odot$).  As will be seen later, no overwhelming metallicity
effects have been found.

With increasing redshift, the rest wavelengths probed by each filter
move more and more blueward (by $z\approx 3$, the $H$ filter is
centered at 4000\AA) and consequently less and less information is
available about the age-sensitive rest-optical part of the SED.  Thus,
past $z=3$, only one filter ($K_s$) is redward of the 4000\AA\ break.
Because of this loss of age-sensitive information at higher redshifts,
we have chosen to split the sample at $z=3$.  We consider fits for the
11 objects at $z<3$ to be much more reliable than those for the six at
$z>3$.  The results and discussion presented below are mainly based on
the 11 $z<3$ galaxies; we will, however, retain the six $z>3$ objects
for completeness and as a consistency check.

\subsection{Dust Content}\label{fit_dust_content}

\begin{figure}
\plotone{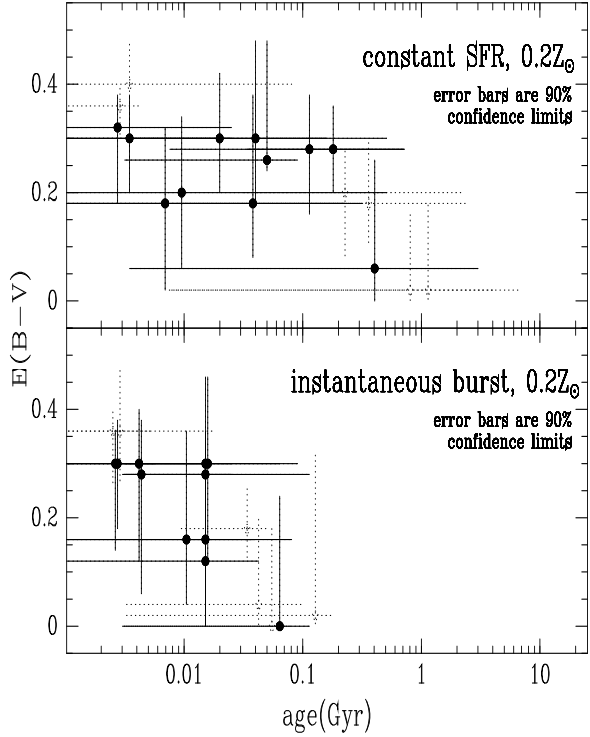}
\figcaption[Sawicki.fig2.ps]
{ \label{Ebv_vs_age.fig} Reddening and age of the best-fit
$Z=0.2Z_{\odot}$ models.  The top panel shows the constant SFR fits
and the bottom one is for the instantaneous burst model.  Age is the
time since the onset of star formation.  Galaxies at $z<3$ are shown
as solid symbols while those at $z>3$, for which fit quality is
poorer, use broken ones.  Error bars come from $\chi^2$ fitting and
correspond to 90\% confidence limits.  $Z=0.02Z_{\odot}$ and
$1.0Z_{\odot}$ models produce similar results.  }
\end{figure}

For each object the models described above fit simultaneously the dust
content and the age of the dominant stellar population.  The age and
$E(B-V)$ parameters of the best-fitting $0.2Z_\odot$ models are
plotted in Figure~\ref{Ebv_vs_age.fig}, with error bars corresponding
to 90\% confidence limits.  The Figure illustrates that the Lyman
break objects are best fitted by models which require non-zero amounts
of dust.  The median values for both the instantaneous burst and
constant SFR $0.2Z_\odot$ scenarios are $E(B-V)=0.28$, producing a
factor of 16 attenuation at 1600\AA.  This attenuation value is in
agreement with the result derived from the UV spectral indices by
Meurer et al.\ (1997) (see also \cite{cal97}).  Note, however, that
Meurer et al.\ had to assume (correctly, as we shall see in
\S~\ref{fit_ages}) that the $z>2$ galaxies are no older than 0.3 Gyr.

\begin{figure}
\plotone{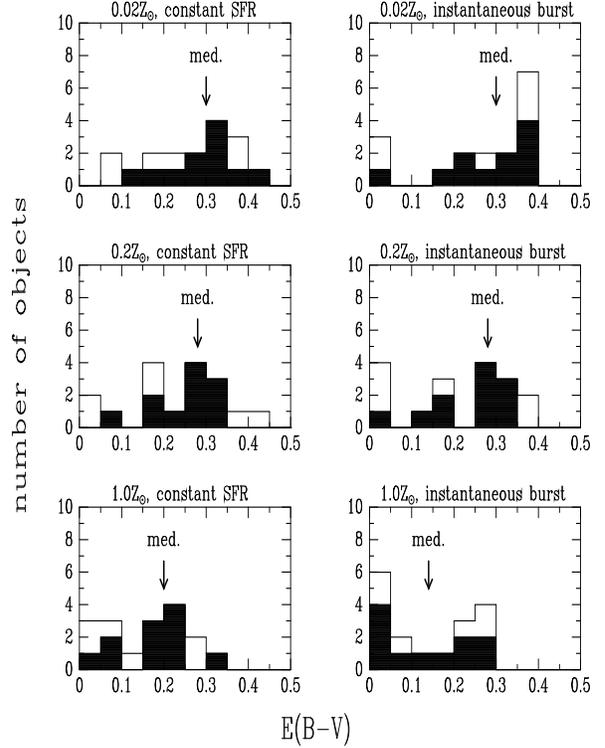}
\figcaption[Sawicki.fig3.ps]
{ \label{Ebv_hist.fig} Intrinsic reddening for the 6 models of
different metallicity and star formation history.  The shaded
histogram corresponds to those galaxies at $z<3$, while the unshaded
one is for all 17 objects (i.e., including the lower-quality $z>3$
fits).  Arrows indicate the median values of $E(B-V)$ for the $z<3$
galaxies.  Substantial amounts of dust are present under all
scenarios. }
\end{figure}

Just as in the $0.2Z_\odot$ case, the $0.02Z_\odot$ and $1.0Z_\odot$
models reqiure significant amounts of dust.  The results are
summarized in Figure~\ref{Ebv_hist.fig} which shows the distribution
of $E(B-V)$ values for the six different scenarios.  Dust is required
for the vast majority of galaxies in all of the six scenarios.  For
sub-solar metallicities ($0.02Z_\odot$ and $0.2Z_\odot$) typical values
are $E(B-V) \approx 0.3$ and somewhat lower ($E(B-V) = 0.15-0.2$) for
the unlikely case of solar metallicity.  

\begin{figure}
\plotone{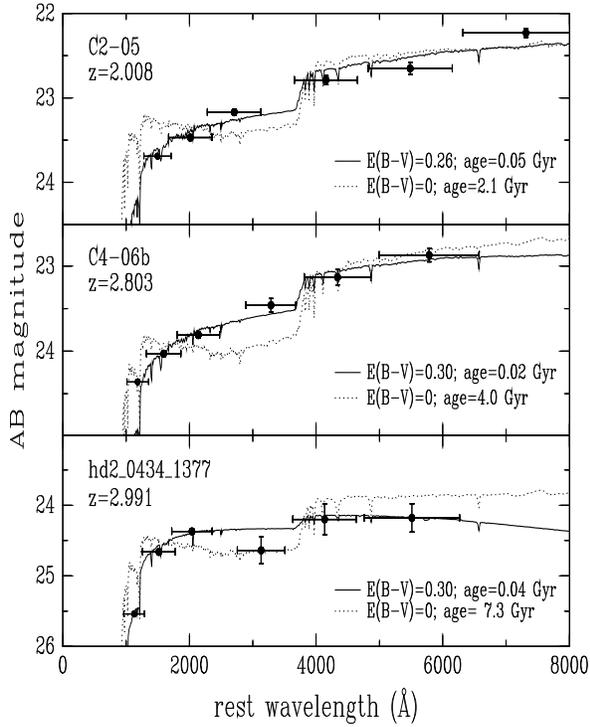}
\figcaption[Sawicki.fig4.ps]
{ \label{noDust_fits.fig} Examples of fits with and without dust for
the $Z=0.2Z_\odot$ constant SFR models.  The broken line shows the
best-fit models for which age was allowed to vary freely, but $E(B-V)$
was held at 0.  The solid line shows the best-fit model (as in
Figure~\ref{best_fit_mosaic.fig}) for which both age and dust were
free parameters.  Models with dust produce better fits than the
dust-free ones. }
\end{figure}

Can dust-free models be securely ruled out?  The error bars in
Figure~\ref{Ebv_vs_age.fig} are 90\% confidence limits; hence, at 90\%
confidence, non-zero amounts of dust are unavoidable in all but one or
two of the 11 $z<3$ objects.  The presence of dust is further
illustrated in Figure~\ref{noDust_fits.fig}, where, for three objects,
the best possible dust-free models are compared with those in which
dust is allowed.  More qualitatively, the median reduced $ \chi ^2$ are
1.34 and 1.68 for the instantaneous burst ad constant SFR $0.2Z_\odot$
fits with dust; the corresponding dust-free values are 5.04 and
10.1. Models with dust give much better fits than those without it.
Therefore we conclude that dust is present in Lyman break galaxies.  A
similar conclusion was reached by Ellingson et al.\ (1996) for the
$z=2.72$ galaxy cB58, which has relatively high-precision IR
photometry.

The high reddening values obtainded here from broadband photometry of
Lyman break galaxies are in contrast to the negligible extinctions
inferred from chemical abundances in damped Ly$\alpha$ systems
(\cite{lu96}; \cite{pet97a}).  However, while quasar sightlines probe
essentially random regions of damped Ly$\alpha$ galaxies, broadband
photometry targets regions of star formation.  Thus, if the
star-forming regions have substantially different dust-to-gas ratios
than ``random'' parts of high-redshift galaxies, then the discrepancy
between the damped Ly$\alpha$ spectroscopic and our photometric
results may simply reflect the differences in the environments that
the two methods probe.

Star formation rates have been derived from the fluxes of
either H$\alpha$ or H$\beta$ for the small number of high-redshift
objects for which these fluxes have been measured (\cite{bec97};
\cite{pet97b}).  In these objects, line-based SFR measurements are 
generally lower than the dust-corrected star-formation rates derived
 from UV fluxes.  This apparent discrepancy can, however, be caused by
 a number of effects other than discrepant dust measurements; these
 effects include leakeage of Lyman-limit photons from
\ion{H}{2} regions, absorption of Lyman-limit photons by dust, and a 
non-standard IMF (see \cite{bec97} and references therein).  These
effects may affect the reliability of SFR measurements based on emission
lines.  Ultimately, {\it concurrent} measurements of
H$\alpha$/H$\beta$ ratios will provide a robust and independent
estimate of the dust content in Lyman break galaxies.

%While emission line measurements can constrain the amount of
%dust, {\it concurrent} measurements of H$\alpha$ {\it and} H$\beta$
%are needed.

Based on our analysis, the presence of dust in Lyman break galaxies
seems unavoidable with $E(B-V) \approx 0.3$ being typical.  As will be
discussed in \S~\ref{discuss_SFR}, these amounts of dust are large
enough to significantly affect estimates of star formation and metal
ejection rates at high redshift.

\subsection{Ages of Stellar Populations}\label{fit_ages}

%The most visible component of the stellar population is the one which
%has undergone the most recent burst of star formation.  
For each galaxy, we shall date the age of the galaxy's dominant
stellar population.  It must be emphashized that the age of the
dominant stellar population does not mean the age of the galaxy
itself, but rather corresponds to the time since the onset of the most
recent major episode of star formation.  Note that the instantaneous
burst and constant SFR models are the limiting-case scenarios, while a
galaxy's star formation history probably lies somewhere between these
two extremes.  Thus, the instantaneous burst model provides the lower
bound, while the constant SFR model gives the upper bound on the
likely age of the galaxy's current episode of star formation.

Figure 2 presents the parameters of the best-fitting $0.2Z_\odot$
models, while Figure 5 shows the distribution of galaxy ages for all
six combinations of metallicity and star formation history.  As these
Figures illustrate, in the constant SFR scenario the vast majority of
$z<3$ objects, for which spectral coverage is sufficient in the red,
are best fit with models for which the ages of dominant stellar
populations are less than 0.2 Gyr.  In the instantaneous burst
scenario, the ages of virtually all the objects, including the six
with $z>3$, are less than 0.1 Gyr.  The median ages for the $z<3$
objects are 36 Myr and 10 Myr in the $0.2Z_\odot$ constant SFR and
instantaneous burst scenarios, respectively.  Although objects at
$z>3$ have insufficient IR coverage to constrain ages reliably, their
fits are also consistent with very young stellar populations; for the
instantaneous burst scenarios these ages are as young as for the $z<3$
objects.

How confidently can we rule out older ages of stellar populations?
The error bars in Figure~\ref{Ebv_vs_age.fig} are 90\% confidence
limits.  If we consider the constant SFR model (which is more
favorable for long ages), we see that of the $z<3$ objects all but one
must have ages $<$~1~Gyr at the 90\% confidence level.  A further
illustration of the necessity of young ages is given in
Figure~\ref{1Gyr_fits.fig}: the (forced) 1~Gyr-old models produce
poorer fits than the unforced younger models.  More qualitatively,
whereas the median reduced $\chi^2$ are 1.34 and 1.68 for the
$0.2Z_\odot$ instantaneous burst and constant SFR models with free
age, the corresponding values for the forced 1Gyr fits are 355.2 and
6.67, respectively.  We therefore conclude that the $z>2$ Lyman break
galaxies are dominated by very young stellar populations.

A possible concern is that the IMF employed to generate the models
(i.e., the Salpeter $0.1< \frac{m}{M_\odot} < 125$ IMF) may not
reflect the true IMF in the Lyman break galaxies.  In particular, the
determination of the age of the stellar population is primarily
sensitive to the size of the Balmer break, which grows as the
population ages.  Since low-mass stars are responsible for the Balmer
break, an IMF deficient in low-mass stars would produce a Balmer break
that is smaller than expected for a given age, thereby mimicking the
signature of a younger stellar population.  Consequently, for an IMF
deficient in low-mass star, our fits would underestimate the ages of
stellar populations.  The multi-metallicity SEDs of Bruzual
\& Charlot (1996) are available only for the $0.1 <
\frac{m}{M_\odot} < 125$ IMF.  Hence, to estimate the impact
of an IMF with a high lower-mass cutoff, we have compared the
predicted colors of Bruzual \& Charlot (1993) solar metallicity models
with different IMF mass ranges.  Specifically, we have compared the
$0.1 < \frac{m}{M_\odot} < 125$ and $2.5 < \frac{m}{M_\odot} < 125$
Salpeter IMFs.  The comparison reveals that, for the young ages in
question and {\it if} the IMF is deficient in low-mass stars, the age
of the star-forming population can be underestimated by a factor of
$\sim 5$ in the instantaneous burst scenarios; in the constant SFR
models the underestimate is only at the 10\% level.  These effects
would then increase the median age in either scenario to no more than
$\sim 50$ Myr.  IMF effects are thus unlikely to affect our conclusion
that the Lyman break objects have very young stellar populations.

Figure~\ref{age_z_2models.fig} shows, as a function of redshift, ages
of stellar populations obtained with the $0.2Z_\odot$ SEDs.  The age
of the $q_0=0.5$, $t_0=12.5$ Gyr Universe and the age of an object
which formed at $z_f=4.5$ are shown for comparison.  Lyman break
galaxies are, by and large, dominated by very young stellar
populations.  The absence of older (age $>0.2$ Gyr) objects,
particularly at the lower end of the redshift range covered, is
remarkable.  Possible reasons for this absence will be discussed in
\S~\ref{discuss_ages}.

\subsection{Star Formation Rates}\label{SFRs}

To estimate the star formation rates in Lyman break galaxies, we use
the constant SFR fits.  For each of the observed objects, we obtain
the SFR by comparing the normalization of the fit to that of a
$1M_\odot yr^{-1}$ model of identical age and reddening.  For the
$z<3$ objects, $0.2Z_\odot$ constant SFR models, and $q_0=0.5$, the
median SFR is $59h_{100}^{-2} M_\odot yr^{-1}$ ($167h_{100}^{-2}
M_\odot yr^{-1}$ for $q_0 = 0.05$).  The distribution of star
formation rates in our sample is shown in
Figure~\ref{SFR.fig}.

As a check on the star formation rates derived above from SED fitting,
we can use the predicted rest-frame 1500~\AA\ flux of a 9 Myr old,
$10^6 M_\odot$ galaxy with a $100< \frac{m}{M_\odot} < 1 $ Salpeter
IMF (\cite{lei95}).  Comparing the dust-corrected, observed fluxes of
the HDF Lyman break objects against the fiducial flux of Leitherer et
al.\ gives a median SFR of $63h_{100}^{-2} M_\odot yr^{-1}$
($q_0=0.5$), thereby confirming the SFR value obtained earlier from
SED fitting.
%Leitherer et al.\ gives a median SFR of $63(179)h_{100}^{-2} M_\odot
%yr^{-1}$ for $q_0=0.5(0.05)$, confirming the SFR value obtained
%earlier by SED fitting.

For illustrative purposes, we have also computed the SFRs under the
assumption that the Lyman break objects are dust-free.  The resulting
median star formation rate is $2h_{100}^{-2} M_\odot yr^{-1}$ for
$q_0=0.5$ ($6h_{100}^{-2} M_\odot yr^{-1}$ for $q_0=0.05$), in
agreement with the dust-free SFR estimates given by Steidel et al.\
(1996a) and Lowenthal et al.\ (1996).  However, since models which
include dust produce much better fits, we prefer the $\sim 60
h_{100}^{-2} M_\odot yr^{-1}$ ($\sim 170 h_{100}^{-2} M_\odot yr^{-1}$
for $q_0=0.05$) dust-corrected rates of star formation.

\subsection{Stellar Masses}\label{masses}

To estimate the stellar masses produced in Lyman break galaxies, we
can use the results of either the instantaneous burst fits or the
constant SFR fits.  For the instantaneous burst fits, we compute the
stellar mass formed in each of the observed objects by comparing the
normalization of the fit to that of a $1M_\odot$ model of identical
age and reddening.  For the $z<3$ objects, $0.2Z_\odot$ instantaneous
burst models, and $q_0=0.5(0.05)$, the median mass in stars is $1(2)
\times 10^9 h_{100}^{-2} M_\odot$, or about $\frac{1}{30}$ ($\frac{1}{15}$)
the stellar mass of a present-day $L^*$ galaxy ($ 3 \times 10^{10}
h_{100}^{-1} M_\odot $; \cite{cow95}).

%Under the constant SFR assumption, the stellar mass produced can be
%estimated by multiplying the SFR by the time since the onset of star
%formation.  

Under the constant SFR assumption, the typical duration of a
starforing espisode is $2 \times$ the median observed age.  Using the
ages and SFRs from the constant SFR fits, we get the median stellar
mass to be $2(4) \times 10^9 h_{100}^{-2} M_\odot$ for $q_0=0.5(0.05)$
--- in agreement with the value derived from instantaneous burst fits.
As was the case for star formation rates, comparing dust-corrected UV
fluxes to Leitherer et al.\ models confirms the above masses.  The
distributions of stellar masses for both the instantaneous burst and
constant SFR models are shown in
Figure~\ref{masses.fig}(b--c).  In both scenarios, the stellar
masses of Lyman break objects are generally smaller than the stellar
mass of a present-day $L^*$ galaxy.  Stellar masses appear to be
relatively insensitive to the details of star formation history.

As an illustration, the above analysis can also be applied to the fits
obtained with dust-free models.  For $q_0=0.5(0.05)$, under the
instantaneous burst assumption, the resulting median stellar mass is
$1(4) \times 10^9 h_{100}^{-2} M_\odot$.
%from SED fitting, and $0.4(1) \times 10^8
%h_{100}^{-2} M_\odot$ from the rest-UV fluxes.  
Using the ages and star formation rates of the constant-SFR fits gives
the median stellar mass of $6(16) \times 10^9 h_{100}^{-2} M_\odot$.
% from SED fitting, and $1(4)
%\times 10^9 h_{100}^{-2} M_\odot$ from the rest-UV fluxes.  
These values, obtained from dust-free model fits, are surprisingly
similar to the stellar masses derived from models which account for
dust.  In the instantaneous burst case, this agreement arises because
the dust-free fits give older ages, which necessitates larger stellar
masses to account for the observed fluxes.  In the constant SFR case
the agreement occurs because the lower star formation rate of the
dust-free models are offset by their older ages.

%The relative insensitivity of total
%stellar mass to the galaxies' dust content gives the robust result
%that typical stellar mass produced by the ongoing episode of 
%star formation in Lyman break objects is $\sim 10^9
%M_\odot$.

%***The median stellar masses (both in the instantaneous burst and
%constant SFR cases), are of the same order of magnitude as those
%derived from the dust-corrected fits.  This agreement arises because
%even though the intrinsic fluxes for the dust-free scenarios are
%lower, the corresponding ages of stellar populations are higher.  This
%insensitivity to dust correction allows the robust conclusion that in
%a typical HDF Lyman break galaxy, the stellar mass produced in the
%most recent episode of star formation is a~few~$\times 10^9
%h_{100}^{-2} M_\odot$.  This stellar mass is a factor 10 to 100
%smaller than that in a present-day $L^*$ galaxy.

The relative insensitivity of the total stellar mass formed to the
assumed star formation history and dust content of the galaxy allows
us to obtain a robust result for the typical mass produced by a star
formation episode in a Lyman break galaxy.  Averaging the median
masses calculated from the dust-corrected instantaneous burst and
constant SFR fits, we have that the median stellar mass produced in an
episode of star formation in an HDF Lyman break galaxy is $\sim 1.5(3)
\times 10^{9} h_{100}^{-2} M_\odot$ for $q_0=0.5(0.05)$.  This median
stellar mass is within the $10^{10}M_\odot$ upper limits on the {\it
gravitational} masses of HDF Lyman break galaxies which were obtained
by Lowenthal et al.\ (1997) on the basis of Ly$\alpha$ emission line
widths and is equivalent to $\sim \frac{1}{20}$ ($\frac{1}{10}$) the
stellar mass of a present-day $L^*$ galaxy.

\section{IMPLICATIONS}\label{implications}

\subsection{Dust Corrections}
%to Star Formation and Metal Production Rates}
\label{discuss_SFR}

Intervening dust has the effect of absorbing UV photons 
%and re-emitting them in the thermal IR, 
,thereby suppressing the observed UV flux.  If, as the SED fits
discussed in \S~\ref{fit_dust_content} indicate, $z>2$ objects have
substantial amounts of internal reddening, recent measurements of the
rates of star formation (\cite{ste96a};
\cite{mad96}; \cite{low97}) and metal ejection (\cite{mad96};
\cite{saw97}) need to be revised.

Steidel et al.\ (1996a) have used rest-1500\AA\ fluxes to estimate the
average star formation rate of their Lyman break galaxy sample to be
$2h_{100}^{-2} M_\odot yr^{-1}$ for $q_0=0.5$ (and $6h_{100}^{-2}
M_\odot yr^{-1}$ for $q_0=0.05$).  If we assume that their objects
suffer from the same amount of reddening as is typical for the Lyman
break objects in the Hubble Deep Field ($E(B-V) \approx 0.28$), then
their SFR estimates need to be adjusted upward to
$38h_{100}^{-2} M_\odot yr^{-1}$ for $q_0=0.5$ (and to $111h_{100}^{-2}
M_\odot yr^{-1}$ for $q_0=0.05$). 

On the basis of 1500\AA\ fluxes, Lowenthal et al.\ (1997) made
estimates of star formation rates both in individual $z \approx 3$ HDF
objects and in the volume-averaged population.  
Reddening correction 
%by a factor of 18 
would bring the Lowenthal et al.\ range of star formation rates to
$13-35h_{100}^{-2} M_\odot yr^{-1}$ for $q_0=0.5$ ($31-111h_{100}^{-2}
M_\odot yr^{-1}$ for $q_0=0.05$).  The volume-averaged star formation
rates given by Lowenthal et al.\ need likewise be adjusted upwards by
a factor of $\sim 18$.
%(It should be
%noted that Lowenthal et al.\ reckoned that their star formation rates
%may underestimate the true SFRs by a factor of 2 or more because of
%the presence of dust).

Madau et al.\ (1996) studied a sample of U- and B-band dropouts in the
Hubble Deep Field.  They used the comoving UV (1620 \AA\ at $\langle z
\rangle = 2.75$ and 1630 \AA\ at $\langle z \rangle = 4.0$) luminosity 
densities to estimate the volume-averaged rates of metal ejection and
star formation.  If we apply the typical $E(B-V)=0.28$ extinction
found to occur in $z>2$ galaxies, the Madau et al.\ estimates of both
the star formation and metal ejection rates need to be revised upward
by a factor of 16.  This value is consistent with the factor of 15
derived by Meurer et al.\ (1997) on the basis of optical data alone
and with the (as we have seen, correct) assumption of young ages.
Note, however, that the Madau et al.\ (1996) estimates do not include
incompleteness corrections, so they remain lower limits even after
having been corrected for dust.

As part of their photometric redshift study of the Hubble Deep Field,
Sawicki et al.\ (1997) have calculated the expected metal density of
the Universe as a function of redshift.  To do so, they have used
rest-3000\AA\ luminosity densities.  Applying $E(B-V)=0.28$ of
extinction yields a factor of 7 increase in flux at 3000\AA,
requireing a factor of 7 increase in the $2 \lesssim z \lesssim 3.5$
metallicity of the Universe shown in Fig.\ 10. of Sawicki et al.

In summary, the values of star formation and metal ejection rates
presented in the literature are underestimates, since they do not
account for intrinsic reddening.  Though the exact correction factors
depend on the wavelength used for the original estimate, once
correction for dust is made, both star formation and metal ejection
rates go up by about an order of magnitude.  Note that although the
total mass of stars formed per object is similar for the dust-free and
dust-included models (\S~\ref{masses}), the difference in the
volume-averaged star formation and metal ejection rates of the
Universe enters via the very different duty cycle of visibility which
the Lyman break objects have in these two models.

\subsection{Ages and Star Formation Histories of Lyman Break Galaxies}
\label{discuss_ages}

For all but one of the $z<3$ HDF Lyman break galaxies, the ages of the
dominant stellar populations are $<0.2$ Gyr.  Since our redshift range
($2<z<3.5$) spans a Gyr in time, the lack of a significant number of
galaxies with old ($>0.2$ Gyr) stellar populations suggests that, in
Lyman break galaxies, the duration of star formation is short rather
than extended and continuous.

In the rest of this section we will speculate on some of the possible
interpretations of this episodic star formation.  Using simple
calculations, we will show that under this single-burst scenario the
number densities of Lyman break objects, in addition to their stellar
masses, are consistent with Lyman break objects being progenitors of
present-day $L<L^*$ galaxies.  They are unlikely to be the present-day
$L \geq L^*$ galaxies caught in the act of creation unless one invokes
more complicated mechanisms such as mergers or recurrent bursts of
star formation.

In the simplest scenario, akin to the low-mass starburst model of
Lowenthal et al.\ (1997), each Lyman break galaxy undergoes a single
episode of star formation.  Since the stellar mass produced in a
typical Lyman break galaxy is $\sim 10^9 h_{100}^{-2} M_\odot$
(\S~\ref{masses}), such a burst would result in a low-mass object
whose stellar mass is $\sim \frac{1}{20}$ that of a present-day $L^*$
galaxy.  This typical mass is thus too low to account
for the whole stellar mass of the bulge of a present-day $L^*$ galaxy.

The young ages of their stellar populations, together with the fact
that they will fade rapidly after star formation ends (by $\sim 1$
magnitude in a mere 30 Myr), mean that only a small fraction of such
short-burst galaxies are visible at any one time.  The redshift range
covered by the HDF Lyman break sample ($2<z<3.5$), spans a time
interval of $\sim 0.8$ Gyr in a $h_{100}=0.7$, $q_0=0.5$ universe
($\sim 2.3$ Gyr for $h_{100}=0.7$, $q_0=0.05$).  The median age from
the $0.2Z_\odot$, constant SFR fits is $\sim 35$ Myr.  Assuming that a
typical star-forming burst lasts twice that time, and that it takes
$\sim 30$ Myr for the galaxy to fade $\sim 1$ magnitude (and hence out
of the spectroscopic sample), we estimate that a typical Lyman break
galaxy stays in the sample for $\sim 100$ Myr.  Therefore, for every
galaxy that is detected, there will be $\sim 8$ ($\sim 23$ for
$q_0=0.05$) objects which have faded beyond spectroscopic
detectibility.  Hence, in the volume and redshift range sampled, there
should be a total of $\sim 136$ ($\sim 391$, for $q_0=0.05$) visible
{\it plus} faded objects which are undergoing, or have undergone,
brief but intense bursts of star formation.
% with each such burst
%producing $\sim 10^9 M_\odot$ of stellar mass.  
%It is possible that
%even these enhanced numbers are underestimates since, as Lowenthal et
%al.\ (1997) point out, a number of Lyman break galaxy candidates in
%the HDF have never been observed due to time constraints.

The comoving number density of this combined (both visible and faded)
population is $7 \times 10^{-2} h_{100}^3 {\mathrm Mpc}^{-3}$
($4\times 10^{-2} h_{100}^3 {\mathrm Mpc}^{-3}$ for $q_0=0.05$).  In
the $K_s$-band (which corresponds to $\sim$ rest-$V$), the HDF Lyman
break galaxies span a range of $\sim 2.5$ magnitudes.  Locally, over a
corresponding magnitude range, one expects to see $\sim 2 \times
10^{-2} h_{100}^{-3} {\mathrm Mpc}^{-3}$ galaxies with $L \approx
\frac{1}{20} L^*$ (\cite{lov92};
\cite{lin96}).  The number density of HDF Lyman break objects is thus
similar to, or perhaps somewhat higher than, that of local
$\frac{1}{20} L^*$ galaxies.  Thus, on the basis of their number
densities and stellar masses, Lyman break objects are consistent with
being progenitors of present-day sub-$L^*$ galaxies.

One could postulate that present-day $L>L^*$ galaxies formed through
bursts of star formation which are analogous to those which seem to be
producing the 
%observed objects with stellar masses of $\sim 10^9
%h_{100}^{-2} M_\odot$, 
above-mentioned low-mass objects, but which are more intense though
less numerous than the bursts seen in the HDF.
%The
%number density of present-day $L>L^*$ galaxies, obtained by
%integrating the bright end of the Lin et al.\ (1996) luminosity
%function, is $4.8 \times 10^{-3} h_{100}^3 {\mathrm Mpc}^{-3}$.  The
%densities of the Lyman break objects (both visible and faded) are an
%order of magnitude higher than that local number, and so they are
%unlikely to be the direct progenitors of present-day $L>L^*$ galaxies.
%Furthermore, 
Based on the present-day luminosity function of Lin et al.~(1996), and
assuming that the 17 objects seen in the HDF are precursors of
present-day sub-$L^*$ galaxies, one would expect the HDF to yield
$\sim 5$ Lyman break objects with stellar masses $\geq 3 \times
10^{10} M_\odot$, the expected stellar mass of an $L^*$ galaxy.  The
fact that such massive stellar populations are absent
(Figure~\ref{masses.fig}) implies that most present-day
massive galaxies did not form in single bursts of star formation at
$z>2$.

There are, however, alternative ways to assemble massive objects.  One
option is through galaxy-galaxy merging: the present-day merger rate
is rather low but increases as $(1+z)^{\sim 3-4}$, at least to
moderate redshift (e.g., \cite{yee95}; \cite{pat97}).  If this
redshift trend continues to $z\sim3$, then the merging of a dozen or
so Lyman break objects can easily result in the spheroid of a
present-day $L^*$ galaxy.

Another possibility is akin to the ``Christmas tree'' model of
Lowenthal et al.\ (1997; see also \cite{col97}).  In this model, star
formation is episodic but recurrent within each galaxy: star-forming
episodes are separated by quiescent intervals during which the galaxy
temporarily fades out of the sample.  The underlying stellar
population, resulting from previous star-forming bursts, would be too
faded to be detectable in the presence of an ongoing burst (e.g.,
\cite{ell96}).  Each of the recurrent star-forming bursts would then add, 
typically, $\sim 10^9 h_{100}^{-2} M_\odot$ of new stars to the
galaxy, gradually building up its stellar mass.  A spheroid containing
$\sim 3 \times 10^{10} h_{100}^{-1} M_\odot$ worth of stars could be thus
accumulated in a dozen or two such recurrent starburst episodes.

%Although it is not clear as to whether it is recurrent or not, star
%formation in Lyman break galaxies is episodic.  The deficit of older,
%continuously star-forming populations has ruled out the option of
%continuous star formation of prolonged duration.

\section{CONCLUSIONS}\label{conclusions}

We have fitted the broadband spectral energy distributions of
spectroscopically confirmed $z>2$ HDF Lyman break objects, using the
multi-metallicity spectral synthesis models of Bruzual \& Charlot
(1996).  In the fits we have included correction for internal dust
extinction typical of local star-forming galaxies (Calzetti 1997).
The fits also assume that the IMF at high redshift is not unlike
that seen locally, although some leeway is allowed, particularly in
the lower mass cutoff.

We find that Lyman break galaxies are dominated by very young stellar
populations ($ < 0.2$ Gyr).  The absence of objects with old stellar
populations, particularly at lower redshifts ($z \gtrsim 2$), implies
that star formation in a typical Lyman break galaxy cannot go on
continuously for a prolonged time.  Instead, star formation must occur
in bursts of short duration ($t<0.2$ Gyr).

A typical Lyman break galaxy is shrouded in enough dust to suppress
its UV flux by a factor of $\sim 16$ at 1600\AA.
Consequently, recent UV-based estimates of the rates of star formation
and metal ejection at high redshift (\cite{ste96a}; \cite{mad96}; \cite{saw97};
\cite{low97}) need to be adjusted upwards by a similar factor.

Star formation rates in Lyman break galaxies are high (median of
$59h^{-2} M_\odot yr^{-1}$ for $q_0=0.5)$.  These star formation rates
typically produce $10^9h^{-2} M_\odot$ of stellar mass during the
lifetime of a star-forming episode; this number is robust as it is
relatively insensitive to the details of star formation history and
dust content.  This median stellar mass is equivalent to
$\frac{1}{15}$--$\frac{1}{20}$ the stellar mass contained in a
present-day $L^*$ galaxy.  Objects with larger stellar masses may be
built up through recurrent episodes of star formation, or can be
assembled through mergers.

A stellar population will fade rapidly after the end of a brief
episode of star formation ($\sim 1$~magnitude at rest-2000\AA\ in the
first 30 Myr).  Because star formation in Lyman break galaxies is of
brief duration, a substantial population of objects which have just
recently faded out of the spectroscopic sample may exist at high
redshift.  We estimate that if a typical Lyman break galaxy stays in
our sample for 0.1 Gyr, then there are 8--23 such faded galaxies for
every visible one.  The number density of the combined visible and
faded population (a few $\times 10^{-2}\,h^{3}_{100}\, {\mathrm
mag}^{-1}\, {\mathrm Mpc}^{-3}$) is comparable to the number density
of present-day sub-$ L^*$ ($M_B \gtrsim -18$) galaxies.

On the basis of the evidence presented above, we conclude that Lyman
break objects are likely direct progenitors of present-day {\bf
sub}-$L^*$ galaxies, or that they may form luminous galaxies through
merging or by repeated episodes of star formation.  They do {\it not}
appear to be steadily star-forming direct progenitors of present-day
massive galaxies.

%The vast majority of the spectroscopically confirmed $z>2$ Lyman break
%galaxies in the Hubble Deep Field are best fit with very young SED
%models which include intrinsic dust.  Typical reddening values are
%$E(B-V) \approx 0.3$.  Presence of dust has important implications
%for estimates of both the star formation and metal ejection rates
%based on UV luminosities: at 1500 \AA, $E(B-V) = 0.28$ will suppress
%the UV flux by a factor of 18.  Consequently, UV-based estimates of
%star formation and metal ejection rates in $z>2$ galaxies need to be
%revised upwards by an order of magnitude.

%The $z>2$ galaxies appear to be dominated by very young stellar
%populations, consistent with having undergone major bursts of star
%formation within the last 0.2 Gyr.  While a continuously star-forming
%galaxy would sustain its luminosity, an instantaneous-burst one will
%fade by $\sim 6$ magnitudes in 0.5 Gyr and hence will no longer be
%detectable.  The apparent absence of older objects from the sample
%suggests that star formation in high-redshift galaxies is episodic
%rather than continuous.  

% ACKNOWLEDGEMENTS 
\acknowledgements
We thank Gabriela Mall\'en-Ornelas for a very thorough reading of an
earlier version of this paper.  We also thank Huan Lin, Bob Abraham,
Gerhardt Meurer, and the anonymous referee for discussions and
comments.  This work was financially supported by NSERC of Canada.

\clearpage

% Now comes the reference list.  In this document, we used \cite to call
% out citations, so we must use \bibitem in the reference list, which
% means we use the LaTeX thebibliography environment.  Please note that
% \begin{thebibliography} is followed by a null argument.  If you forget
% this, mayhem ensues, and LaTeX will say "Perhaps a missing item?" when
% you run it.  Do not call us, do not send mail when this happens.  Put
% the silly {} after the \begin{thebibliography}.
%
% Each reference has a \bibitem command to define the citation format
% to be placed in the text (in []) and the symbolic tag used for 
% cross referencing (in {}).
%
% See sample1.tex, or the AASTeX guide, for an alternative to the \cite-
% \bibitem command.

\clearpage

\begin{deluxetable}{lccccccccr}
\tablewidth{0pt}
\tablenum{1}
%\footnotesize
\scriptsize
\tablecaption{Photometry \label{tbl-1}}
\tablehead{
%\colhead{pppn}   & 
\colhead{object}   & 
\colhead{ref}   &
\colhead{$z$}   & 
\colhead{$U_{300,AB}$}   & 
\colhead{$B_{450,AB}$}   & 
\colhead{$V_{606,AB}$}   & 
\colhead{$I_{814,AB}$}   & 
\colhead{$J_{AB}$}   & 
\colhead{$H_{AB}$}   & 
\colhead{$K_{s,AB}$}   
} 
\startdata
C2-05 & S & 2.008 & $25.72 (06)$ & $23.69 (03)$ & $23.47 (03)$ & $23.17 (03)$ & $22.79 (05)$ & $22.65 (07)$ & $22.23 (05)$ 
\nl
hd2\_0725\_1818 & L & 2.233 & $25.83 (05)$ & $24.39 (03)$ & $24.28 (03)$ & $24.19 (03)$ & $24.60 (24)$ & $23.73 (17)$ & $23.69 (16)$ 
\nl
hd2\_2030\_0287 & L & 2.267 & $26.46 (06)$ & $24.47 (03)$ & $24.35 (03)$ & $24.35 (04)$ & $24.16 (15)$ & $24.70 (46)$ & $24.22 (30) $ 
\nl
hd2\_0624\_0266 & L & 2.419 & $28.09 (29)$ & $25.14 (03)$ & $25.06 (03)$ & $25.09 (04)$ & $24.85 (21)$ & $ 24.55(35)$ & $24.52 (31)$ 
\nl
C4-08 & S & 2.591 & $28.62 (46)$ & $24.57 (03)$ & $24.50 (03)$ & $24.31 (03)$ & $24.48 (17)$ & $24.50 (32)$ & $24.17 (22)$ 
\nl
C3-02 & S & 2.775 & $25.84 (05)$ & $24.64 (03)$ & $24.54 (03)$ & $24.50 (03)$ & $24.09 (13)$ & $> 24.27$ & $24.33 (25)$ 
\nl
C4-06a & S\tablenotemark{a} & 2.803 & $27.67 (22)$ & $24.59 (03)$ & $23.47 (03)$ & $23.08 (03)$ & $22.97 (07)$ & $21.96 (05)$ & $21.82 (04)$ 
\nl
C4-06b & S\tablenotemark{a} & 2.803 & $27.33 (12)$ & $24.36 (03)$ & $24.03 (03)$ & $23.81 (03)$ & $23.46 (08)$ & $23.13 (09)$ & $22.87 (08)$ 
\nl
hd4\_0367\_0266 & L & 2.931 & $25.45 (08)$ & $24.77 (04)$ & $23.87 (03)$ & $23.54 (03)$ & $23.73 (18)$ & $22.54 (11)$ & $22.39 (09)$ 
\nl
hd4\_2030\_0851 & L & 2.980 & $ > 29.37 $ & $25.39 (04)$ & $24.74 (04)$ & $24.45 (04)$ & $23.97 (15)$ & $23.97 (27)$ & $23.35 (12)$ 
\nl
hd2\_0434\_1377 & L & 2.991 & $29.74 (26)$ & $25.54 (04)$ & $24.66 (03)$ & $24.37 (03)$ & $24.64 (19)$ & $24.20 (22)$ & $24.18 (20)$ 
\nl
hd2\_1410\_0259 & L & 3.160 & $29.97 (90)$ & $25.38 (04)$ & $24.72 (03)$ & $24.66 (04)$ & $26.02 (97)$ & $24.58 (56)$ & $23.83 (20)$ 
\nl
hd2\_1359\_1816 & L & 3.181 & $ >29.56 $ & $ 25.82 (04)$ & $24.76 (03)$ & $24.35 (03)$ & $24.25 (15)$ & $23.73 (15)$ & $24.47 (26)$  
\nl
C4-09 & S & 3.226 & $26.81 (19)$ & $25.01 (03)$ & $24.12 (03)$ & $23.87 (03)$ & $23.60 (09)$ & $24.01 (24)$ & $ 22.67 (07)$ 
\nl
hd3\_0408\_0684 & L & 3.233 & $ > 29.26 $ & $25.88 (04)$ & $24.89 (04)$ & $24.40 (04)$ & $25.15 (63)$ & $24.98 (78)$ & $ 23.47 (17)$ 
\nl
hd2\_0705\_1366 & L & 3.368 & $27.55 (21)$ & $26.05 (04)$ & $25.04 (03)$ & $24.95 (03)$ & $26.08 (74)$ & $24.24 (21)$ & $24.47 (23)$ 
\nl
hd2\_0698\_1297 & L & 3.430 & $ >29.28 $ & $26.55 (05)$ & $25.19 (03)$ & $24.81 (03)$ & $25.28 (32)$ & $24.93 (45)$ & $ 23.62 (12)$ 
\nl
\tablenotetext{a}{Since object C4-06 has two very distinct components, we 
have treated it as two objects at the same redshift. C4-06a is the lower
object in the C4-06 panel of Fig.\ 2 of Steidel et al.\ (1996b), while 
C4-06b is the upper object.}  
\tablecomments{Upper limits ($2\sigma$) are indicated with a ``$>$''.  Errors
are indicated in parentheses: they are $1\sigma$ uncertainties
expressed as percentages of the {\it flux}.}
\tablerefs{(S) Steidel et al.\ (1996b); (L) Lowenthal et al.\ (1997)}
\enddata
\end{deluxetable}

% figures 

\newpage

\twocolumn

\begin{figure}
\plotone{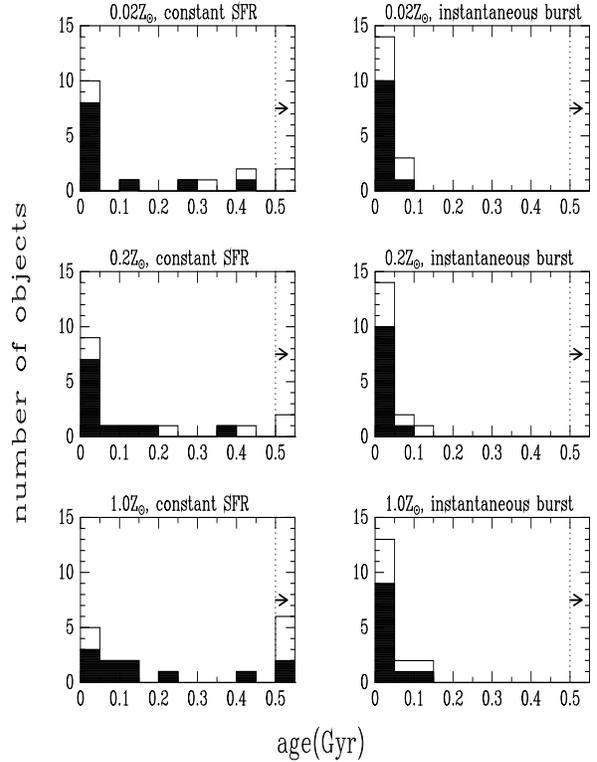}
\figcaption[Sawicki.fig5.ps]
{ \label{age_hist.fig} Ages since the onset of the current episodes of
star formation.  The shaded histogram corresponds to those galaxies at
$z<3$, while the unshaded one is for all 17 objects.  Ages $>0.5$ Gyr
are grouped together in the right-most bins (separated from the other
bins by the dotted lines).  Most Lyman break galaxies are dominated by
very young stellar populations.  }
\end{figure}

\begin{figure}
\plotone{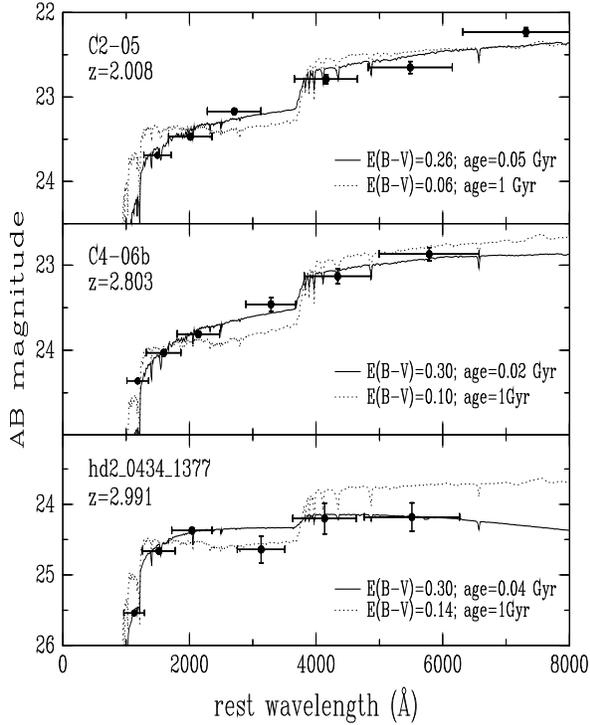}
\figcaption[Sawicki.fig6.ps]
{ \label{1Gyr_fits.fig} Examples of fits with and without forced age
for the $Z=0.2Z_\odot$ constant SFR models.  The broken lines are the
fits for which age was fixed at 1 Gyr (while reddening was allowed to
vary freely).  The solid lines are the fits where, as in
Figure~\ref{best_fit_mosaic.fig}, both age and reddening were free
parameters.  Instantaneous burst models of 1 Gyr age would produce an
even stronger disagreement with the data, since in those models the
UV-producing massive stars would not have continued to be replenished
as in the constant SFR models.  }
\end{figure}

\begin{figure}
\plotone{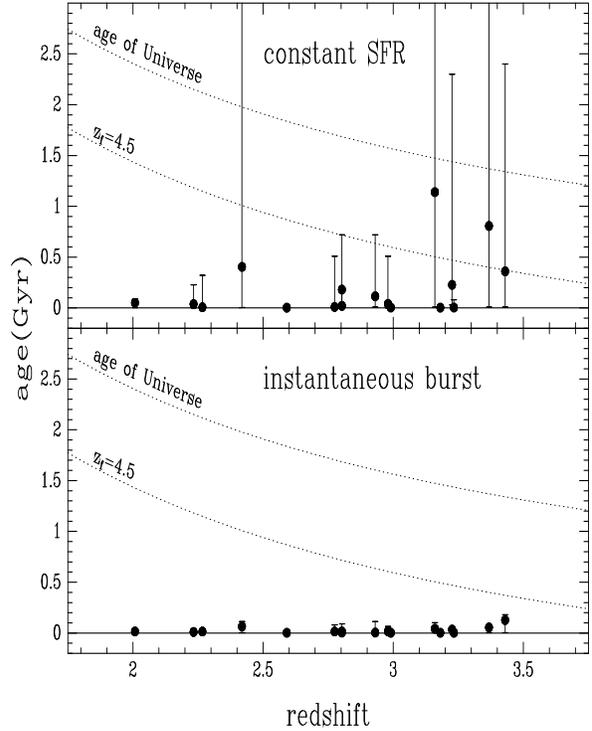}
\figcaption[Sawicki.fig7.ps]
{ \label{age_z_2models.fig} The ages of dominant stellar populations
of Lyman break galaxies determined by fitting $0.2Z_\odot$ model SEDs.
Error bars correspond to 90\% confidence limits.  For comparison, the
dotted lines show the age of the universe and the age of an object
which formed at $z_f=4.5$; both are for a $\Omega=1$ universe with a
present-day age fixed at $t_0=12.5$ Gyr; note, however, that the
stellar population ages do not depend on the value of $H_0$.  Objects
at $z>3$ suffer from poor coverage above $\sim 4000$~\AA\ and,
consequently, have age estimates of lower quality.  Models with
$Z=0.02Z_\odot$ and $1.0Z_\odot$ produce ages similar to these shown
here.  Stellar populations of the majority of $z>2$ HDF galaxies
appear to have undergone recent ($t<0.2$ Gyr) episodes of star
formation.}
\end{figure}

\begin{figure}
\plotone{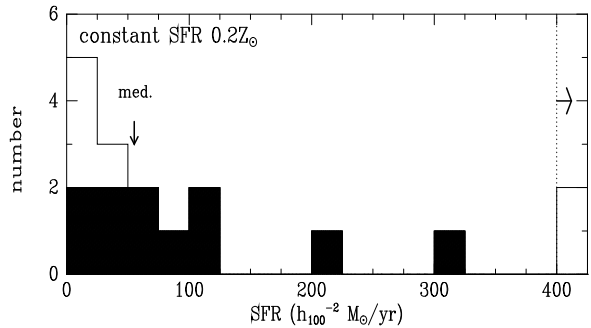}
\figcaption[Sawicki.fig8.ps] 
{ \label{SFR.fig} Distribution of star formation rates obtained from
the $0.2Z_\odot$ constant SFR fits.  A $q_0=0.5$ universe was assumed.
The shaded histogram is for the eleven $z<3$ galaxies, and the
unshaded one includes all 17 objects.  The median value for the $z<3$
subset is indicated.  }
\end{figure}

\begin{figure}
\plotone{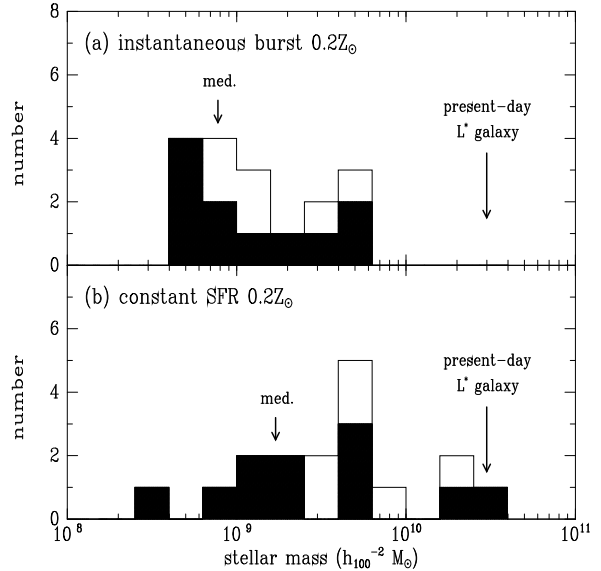}
\figcaption[Sawicki.fig9.ps] 
{ \label{masses.fig} Stellar masses from the $0.2Z_\odot$ fits.
Shaded histograms are for $z<3$ galaxies, and the unshaded ones
include all 17 objects.  Arrows indicate median values for the $z<3$
subset.  A $q_0=0.5$ universe was assumed.  Panel (a) shows the
distribution of stellar masses obtained from the instantaneous burst
fits.  The stellar mass of a present-day $L^*$ galaxy ($3 \times
10^{10} h_{100}^{-1} M_\odot$; Cowie et al., 1995) is indicated for
comparison.  Panel (b) shows the distribution of stellar masses
derived by multiplying the objects' star formation rates and the ages
of their stellar populations. Under both the instantaneous burst and
constant SFR scenarios, stellar masses produced by the observed
episodes of star formation are generally substantially less than the
stellar mass of a present-day $L^*$ galaxy.  }
\end{figure}


\begin{thebibliography}{}

\bibitem[Bechtold et al.\ 1997]{bec97} Bechtold, J., Yee, H.K.C., Elton, R., 
\& Ellingson, E. 1997, \apjl, 477, L29

\bibitem[Bouchet et al.\ 1985]{bou85} Bouchet, P., Lequeux, J., Maurice, E., 
Pr\'evot, L., \& Pr\'evot-Burnichon, M. L. 1985, \aap, 149, 330

\bibitem[Bruzual \& Charlot 1993]{bru93} Bruzual A., G., \& Charlot, S. 
1993, \apj, 405, 538

\bibitem[Bruzual \& Charlot 1996]{bru96} Bruzual A., G., \& Charlot, S. 
1996, in preparation

\bibitem[Calzetti 1997]{cal97} Calzetti D., 1997,  to appear in the 
Proceedings of the Conference ``The Ultraviolet Universe at Low and
High Redshift'' preprint: astro-ph/9706121

\bibitem[Calzetti, Kinney, \& Storchi-Bergmann 1994]{cal94} Calzetti, D., 
Kinney, A., \& Storchi-Bergmann, T. 1994, \apj, 429, 582

\bibitem[Cohen et al.\ 1996]{coh96} Cohen, J. G., Cowie, L. L., Hogg, D. W., 
Songaila, A., Blandford, R., Hu, E. M., \& Snopbell, P. 1996, \apjl,
471, L5

\bibitem[Colley et al.\ 1996]{col97} Colley, W.N., Gnedin, O.Y., 
Ostriker, J.P., \& Rhoads, J.E 1997, \apj, 488, 579

\bibitem[Connolly et al.\ 1997]{con97} Connolly, A.J., Szalay, A.S., 
Dickinson, M., SubbaRao, M.U., \& Brunner, R.J. 1997, \apjl, 486, L11

\bibitem[Cowie et al.\ 1995]{cow95} Cowie, L.C., Hu, E.M., \& Songaila, A.
1995, \nat, 377, 603

\bibitem[Dickinson 1996]{dic96} Dickinson, M. 1996, \\
{\texttt{http://www.stsci.edu/ftp/science/hdf/clearinghouse/irim/irim\_zeropts.html}}

\bibitem[Dickinson et al.\ 1997]{dic97} Dickinson, M., et al.\ 1997, 
in preparation.

\bibitem[Ellingson et al.\ 1996]{ell96} Ellingson, E., Yee, H.K.C., 
Bechtold, J., \& Elston, R. 1996, \apjl, 466, L71

\bibitem[Ferguson 1996]{fer96} Ferguson, H. 1996, 
{\texttt{http://www.stsci.edu/ftp/observer/hdf/logs/zeropoints.txt}}

\bibitem[Fitzpatrick 1986]{fit86} Fitzpatrick, E.L., 1986, \aj, 92, 1068

\bibitem[Fomalont et al.\ 1997]{fom97} Fomalont, E. B., Kellermann, K. I., 
Richards, E. A., Windhorst, R. A., \& Partridge, B. P. 1997, \apjl, 475, L5

%\bibitem[Franx et al.\ 1997]{fra97} Franx, M., Illingworth, G.D., Kelson, D., 
%van Dokkum, P.G., Tran, K.-V. 1997, preprint astro-ph/9704090

\bibitem[Giavalisco et al.\ 1996]{gia96} Giavalisco, M., Steidel, C.C., \&
Macchetto, F.D. 1996, \apj, 470, 189

\bibitem[Gordon, Calzetti, \& Witt 1997]{gor97} Gordon, K. D., Calzetti, D., 
\& Witt, A. N. 1997, \apj, 487, 626

\bibitem[Hogg et al.\ 1997]{hog97} Hogg, D.W., Neugebauer, G., Armus, L., 
Matthews, K., Pahre, M.A., Soifer, B.T., \& Weinberger, A.J. 1997,
\aj, 113, 474

\bibitem[Leither et al.\ 1995]{lei95} Leitherer, C., Robert, C., \& 
Heckman, T.M. 1995, \apjs, 99, 173

\bibitem[Lin et al.\ 1996]{lin96} Lin, H., Kirshner, R.P., Shectman, S.A., 
Landy, S.D., Oemler, A., Tucker, D., \& Schechter, P.L. 1996, \apj,
464, 60

\bibitem[Loveday et al.\ 1992]{lov92} Loveday, J., Peterson, B.A., 
Efstathiou, G., \& Maddox, S.J. 1992, \apj, 390, 338

\bibitem[Lowenthal et al.\ 1997]{low97} Lowenthal, J.D., Koo, D.C., 
Guzm\'an, R., Gallego, J., Phillips, A.C., Vogt, N.P., Illingworth, G.D.,
\& Gronwall, C. 1997, \apj, 481, 673

\bibitem[Lu et al.\ 1996]{lu96} Lu, L., Sargent, W.L.W., Barlow, T.A., 
Churchill, C.W., \& Vogt, S.S. 1996, \apjs, 107, 475

\bibitem[Madau 1995]{mad95} Madau, P. 1995, \apj, 441, 18

\bibitem[Madau et al.\ 1996]{mad96} Madau, P., Ferguson, H.C., 
Dickinson, M.E., Giavalisco, M., Steidel, C.C., \& Fruchter, A. 1996,
\mnras, 283, 1388

\bibitem[Massey et al.\ 1995]{mas95} Massey, P., Lang, C.C., 
DeGioia-Eastwood, K., \& Garmany, C.D. 1995, \apj, 438, 188

\bibitem[Meurer et al.\ 1997]{meu97} Meurer, G.R., Heckman, T.M., 
Lehnert, M.D., Leitherer, C., \& Lowenthal, J. 1997, \aj, 114, 54

\bibitem[Pascarelle et al. 1996]{pas96} Pascarelle, S. M., Windhorst, R. A.,
Keel, W. C., \& Odewahn, S. C. 1996, \nat, 383, 45 

\bibitem[Patton et al.\ 1997]{pat97} Patton, D.R., Pritchet, C.J., 
Yee, H.K.C., Ellingson, E., \& Carlberg, R.G. 1997, \apj, 475, 29

\bibitem[Pettini et al.\ 1994]{pet94} Pettini, M., Smith, L.J., 
Hunstead, R.W., \& King, D.L. 1994, \apj, 426, 79

\bibitem[Pettini et al.\ 1997a]{pet97a} Pettini, M., King, D.L., Smith, L.J., 
\& Hunstead, R.W. 1997, \apj,  478, 536 

\bibitem[Pettini et al.\ 1997b]{pet97b} Pettini, M., Steidel, C.C., 
Adelberger, K.L., Kellogg, M., Dickinson, M., \& Giavalisco, M. 1997,
to appear in Origins, ed. J.M. Shull, C.E. Woodward, and H. Thronson
(ASP Conference Series); preprint: astro-ph/9708117

\bibitem[Sawicki et al.\ 1997]{saw97} Sawicki, M.J., Lin, H., 
\& Yee, H.K.C. 1997, \aj, 113, 1

\bibitem[Serjeant et al.\ 1997]{ser97} Serjeant, S., Eaton, N., Oliver, S., 
Efstathiou, A., Goldschmidt, P., Mann, R.G., Mobasher, B.,
Rowan-Robinson, M., Sumner, T., Danese, L., Elbaz, D., Franceschini,
A., Egami, E., Kontizas, M., Lawrence, A., McMahon, R.,
Norgaard-Nielsen, H.U., Perez-Fournon, I., \& Gonzalez-Serrano,
I. 1997, \mnras, 289, 457

%\bibitem[Songaila et al.\ 1990]{son90} Songaila, A., Cowie, L.L., \& Lilly, 
%S.J. 1990, \apj, 348, 371

%\bibitem[Spinrad et al.\ 1997]{spi97} Spinrad, H., Dey, A., Stern, D., 
%Dunlop, J., Peacock, J., Jimenez, R., \& Windhorst, R. 1997, to appear
%in \apj

\bibitem[Stasi\'nska \& Leitherer 1996]{sta96} Stasi\'nska, G., \& Leitherer,
C. 1996, \apjs, 107, 472

\bibitem[Steidel et al.\ 1996a]{ste96a} Steidel, C. C., Giavalisco, M., 
Pettini, M., Dickinson, M., \& Adelberger, K.L. 1996a, \apjl, 462, L17

\bibitem[Steidel et al.\ 1996b]{ste96b} Steidel, C. C., Giavalisco, M., 
Dickinson, M., \& Adelberger, K.L. 1996b, \aj, 112, 352

\bibitem[Steidel et al.\ 1997]{ste97} Steidel, C. C., Adelberger, K. L., 
Dickinson, M., Giavalisco, M., Pettini, M., Kellogg, M. 1997, \apj, in
press, preprint: astro-ph/9708125

%\bibitem[Trager et al.\ 1997]{tra97} Trager, S.C., Faber, S.M., Dressler, 
%A., \& Oemler, A. 1997, \apj, in press

\bibitem [Wemstaker 1981]{web91} Wemstaker, W. 1981, \aap, 97, 329

\bibitem[Williams et al.\ 1996]{wil96} Williams, R. E., Blacker, B., 
Dickinson, M., Van Dyke Dixon, W., Ferguson, H. C., Fruchter, A. S., 
Giavalisco, M., Gilliland, R. L., Heyer, I., Katsanis, R., Levay, Z., 
Lucas, R. A., McElroy, D. B., Petro, L., Postman, M., Adorf, H.-M., 
\& Hook, R. N. 1996, \aj, 112, 1335

\bibitem [Worthey 1994]{wor94} Worthey, G. 1994, \apjs, 95, 107

\bibitem[Yee 1991]{yee91} Yee, H.K.C. 1991, \pasp, 103, 396

\bibitem[Yee \& Ellingson 1995]{yee95} Yee, H.K.C., \& Ellingson, E.
1995 \apj, 445, 37

\bibitem[Yee et al.\ 1996]{yee96} Yee, H.K.C., Ellingson, E., Bechtold, J., 
Carlberg, R. G., \& Cuillandre, J.-C. 1996 \aj, 111, 1783


\bibitem[Zepf et al.\ 1997]{zep97} Zepf, S.E., Moustakas, L.A., \& 
Davis, M. 1997, \apjl, 474, L1

\end{thebibliography}
\end{document}